\documentclass[aps,prl,twocolumn,floats,showpacs]{revtex4}
\usepackage{graphicx}
\usepackage{amsmath}
\usepackage{amssymb}

\newcommand{\pipulse}{{$\pi$-pulse }}

\begin{document}
\title{Complete transfer of populations from a single state
to a pre-selected superposition of states using
Piecewise Adiabatic Passage: Experiment.}

\date{\today}
\author{S.~Zhdanovich$^{1,3}$, E.A.~Shapiro$^2$, J.W.~Hepburn$^{1,2,3}$, M.~Shapiro$^{1,2,3}$, and V.~ Milner$^{1,2,3}$}
\affiliation{Departments of  Physics \& Astronomy$^1$ and Chemistry$^2$, and The Laboratory for Advanced Spectroscopy and Imaging Research (LASIR)$^3$, The University of British Columbia, Vancouver, Canada}

\begin{abstract}{We demonstrate a method of adiabatic population transfer from a single quantum state into a coherent  superposition of states. The transfer is executed with femtosecond pulses, spectrally shaped in simple and intuitive manner, which does not require iterative feedback-controlled loops. In contrast to non-adiabatic methods of excitation, our approach is not sensitive to the exact value of laser intensity. We show that the population transfer is complete, and analyze the possibility of controlling the relative phases and amplitudes of the excited eigenstates. We discuss the limitations of the proposed control methods due to the dynamic level shifts and suggest ways of reducing their influence.}
\end{abstract}


\maketitle

\section{Introduction}
Population transfer from one energy state to a coherent superposition of states (i.e. a ``wavepacket'') is an important tool in atomic and molecular physics and chemistry. Preparation of a wavepacket with well defined amplitudes and phases of the constituent eigenstates is the starting point for many techniques in precision spectroscopy \cite{Dhar1994, Stolow1998}, coherent control of molecular dynamics and chemical reactions \cite{ShapiroBrumerBook}, in the design of atomic clocks \cite{Diddams2004} and fault tolerant quantum computing schemes \cite{NielsenChuangBook}. In many situations, it is desirable to make the population transfer complete, that is to move the entire ensemble of atoms or molecules to the target superposition state, in a robust manner with respect to the excitation field parameters.

Both adiabatic and  non-adiabatic methods of complete population transfer between two \textit{single} eigenstates are well established \cite{AllenEberlyBook}. Non-adiabatic Rabi cycling in a two-level system enables complete population transfer by means of a so-called ``\pipulse''. Though a \pipulse can be realized on a very short time scale, it is far from being robust, as it is highly sensitive to the exact value of the laser intensity, frequency and pulse length.  The method can be extended to the case of a multi-level target state \cite{IJQCpaper}, but suffers from the following complication.

\begin{figure}
\centering
\includegraphics[width=.8\columnwidth]{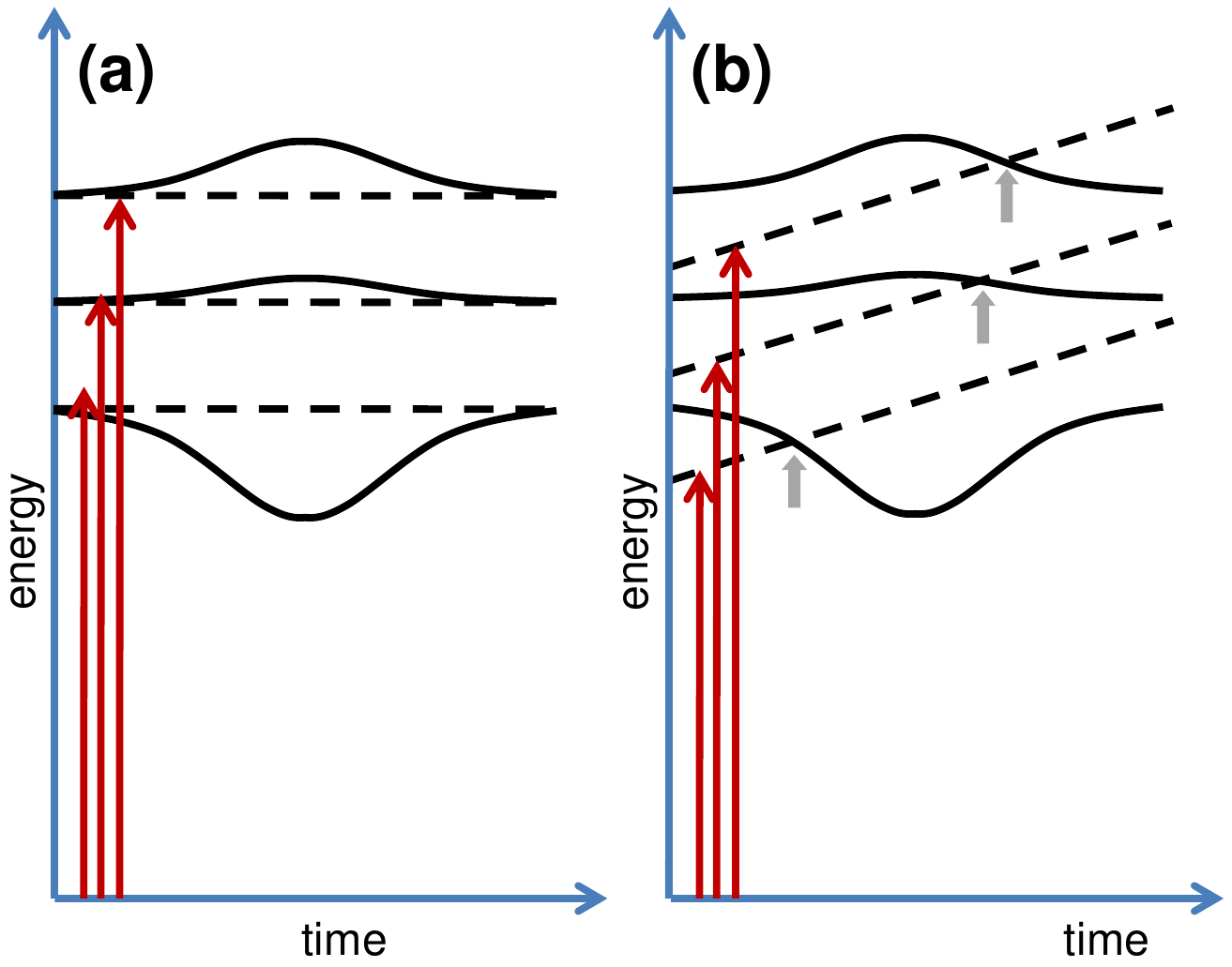}
\caption{(Color online). Illustration of the interaction of a multilevel system with polychromatic coherent radiation consisting of a number of discrete spectral components (red vertical arrows). Dashed lines represent the ground state ``dressed" with a photon. Bare excited states of the system (solid lines) are dynamically shifted in energy due to the presence of non-resonant components of the laser field. This results in incomplete population transfer via non-adiabatic process (\textbf{a}). When the spectral components are simultaneously chirped in frequency (\textbf{b}), adiabatic transfer into a coherent superposition of excited states is executed. Different timing of the level crossings (gray arrows) complicates  the control over the resulting wavepacket, but does not affect the completeness of excitation.}
\label{Fig-LevelDiagram}
\end{figure}

Because of the difference in transition dipole moments to different excited eigenstates, a single transform-limited pulse can not, in general, serve as a \pipulse  for all transitions simultaneously. Recently, we have shown that such non-uniformity of dipole moments can be compensated using the technique of spectral pulse shaping \cite{IJQCpaper}. Rabi frequencies, and therefore pulse areas, for each transition can be equalized by adjusting the amplitudes of the corresponding frequency components of the excitation pulse. The required amplitudes are, however, not easy to find due to the dynamic Stark shifts of the energy levels. As illustrated in Fig.\ref{Fig-LevelDiagram}(a), strong polychromatic field changes the instantaneous resonant frequencies of atomic transitions, making the necessary (for a \pipulse) resonance condition hard to satisfy throughout the whole excitation process. Similar scenario has been considered in a two-photon excitation scheme, where the required pulse shaping has been found by means of iterative feedback controlled loops \cite{Trallero-Herrero2005}.

In contrast to the above, the method of  Adiabatic Passage (AP) is less sensitive to the parameters of the driving field, transition dipole moments and resonant frequencies \cite{Gaubatz90, Vitanov2001, Malinovsky01}. In AP, the dynamics of a quantum system is described by a time-dependent Hamiltonian, whose instantaneous ``adiabatic'' eigenstates are essentially decoupled from one another as long as the field is strong and its parameters are changing slowly \cite{MessiahBook}. Each adiabatic state is a coherent superposition of bare eigenstates of a field-free Hamiltonian. As the parameters of the interaction field change in time, projections of each adiabatic state on the bare-state basis set are evolving accordingly. By making one of the adiabatic states coincide with the initial state of an atom at the beginning of the interaction, and with the target excited state - at the end, one can achieve robust and complete transfer of population as long as the adiabaticity conditions are satisfied throughout the interaction.

Adiabatic passage between two states is most commonly executed by sweeping, or ``chirping'', an instantaneous frequency of the driving field across a resonance. Adiabatic crossing of the excited state by the ground state dressed with a photon field results in complete population transfer between those \textit{two} states.  AP into multi-level wavepackets have been recently considered in two distinct interaction scenarios: with multiple phase-locked narrowband laser fields \cite{Thanopulos2006, Kral2007, Vitanov-transforms}, and a single spectrally shaped broadband pulse \cite{Shapiro2009}. The latter approach provides a simple strategy of shaping the pulse in accord with the energy level structure of the system of interest. Introducing local frequency chirp around each resonant transition frequency (Fig.\ref{Fig-LevelDiagram}(b)), {simultaneous} AP is initiated into multiple target states which make up the final wavepacket. Figure \ref{Fig-LevelDiagram} (b) provides an intuitive illustration of the robustness of the method, not only with respect to the field amplitude and frequency, but also with respect to the dynamic level shifts associated with the presence of a strong polychromatic field. Indeed, crossings of the dressed ground state with multiple excited states will occur even if the latter are heavily perturbed by the off-resonance components of the driving field. Dynamic shifts of transition frequencies result in a slightly different timing of each adiabatic crossing. As we show here, this lack of synchronism between multiple APs slightly complicates the control of the makeup of the target wavepacket, but does not reduce the efficiency of population transfer.

\begin{figure}
\centering
\includegraphics[width=1\columnwidth]{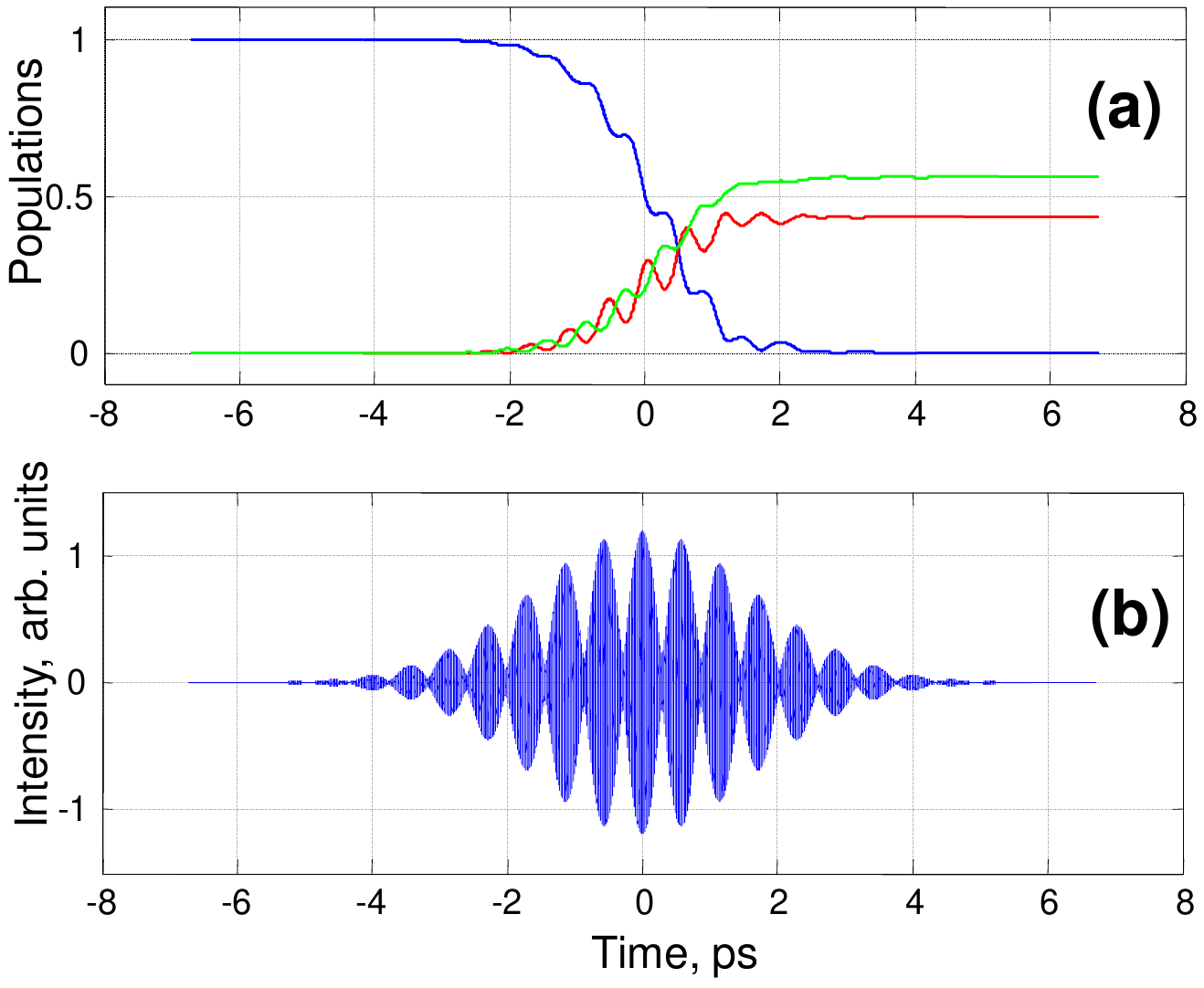}
\caption{(Color online). Results of numerical simulations of piecewise adiabatic passage for Potassium atom. Population transfer is shown in (\textbf{a}) as a function of time. Blue line represents the ground state population while red and green lines correspond to the populations of $4P_{1/2}$ and $4P_{3/2}$ excited states, respectively. Excitation field amplitude is plotted in (\textbf{b}) on the same time scale.}
\label{Fig-APSimulations}
\end{figure}

In the time domain, the interaction picture is especially intuitive when the target states are equidistant in energy. In that case, shaping a pulse with multiple local frequency chirps results in a train of mutually coherent ultrashort pulses, separated by the evolution period of the wavepacket. Though each pulse in the train transfers only a small amount of population to the target superposition state, population of that state coherently accumulates \cite{Dhar1994, CoherentAccumulation1, CoherentAccumulation3} piece by piece, reaching 100\% at the end of the interaction regardless of the total energy of the pulse train \cite{Shapiro2009}. Shown in Fig.\ref{Fig-APSimulations} is an example of numerically calculated dynamics of such Piecewise Adiabatic Passage (PAP) into a wavepacket consisting of only two levels and driven by a train of ultrashort pulses.

The population transfer technique we are studying here can be viewed as a practical application of the $1+N$-level control schemes, based on the Morris-Shore transform \cite{MStransform} and discussed in \cite{Vitanov-transforms}. The selectivity of the transfer is obtained by tailoring the temporal and spectral profiles of the train of pulses to the target wavepacket dynamics \cite{Dhar1994, Walmsley-deAraujo, Wollenhaupt06, deAraujo08}. The correspondence between the properties of the pulse train and the target wavepacket dynamics has also been noted in the optimization studies aimed at either maximizing population transfer into a wavepacket \cite{Grafe05, Trallero-Herrero06, Hertel-JPB08} or stabilizing such transfer against the wavepacket spreading and decoherence \cite{Walmsley-Science08}. Our method provides an alternative to the ``multi-RAP'' pulse sequences of \cite{Baumert-landscapes-08} and ''molecular $\pi$-pulses'' used in exciting molecular wavepackets \cite{MolPiPulse1, MolPiPulse2, MolPiPulse3, MolPiPulse4}. The difference with the latter methods is manifest when the target wavepacket consists of more than two eigenstates \cite{Shapiro2009}.

Recently, we have demonstrated experimentally the method of piecewise adiabatic passage into a single excited state \cite{Zhdanovich2008}. Here, we extend this method to the multi-level target states case, by exciting the fine-structure doublet in atomic Potassium. We show {\it\small experimentally} that one can achieve complete population transfer into such ``spin-orbit'' wavepacket with high degree of robustness. We also present a simple and intuitive approach to controlling the phases and amplitudes of the constituent eigenstates of the final superposition, and address its limitations.
\begin{figure}
\centering
\includegraphics[width=1\columnwidth]{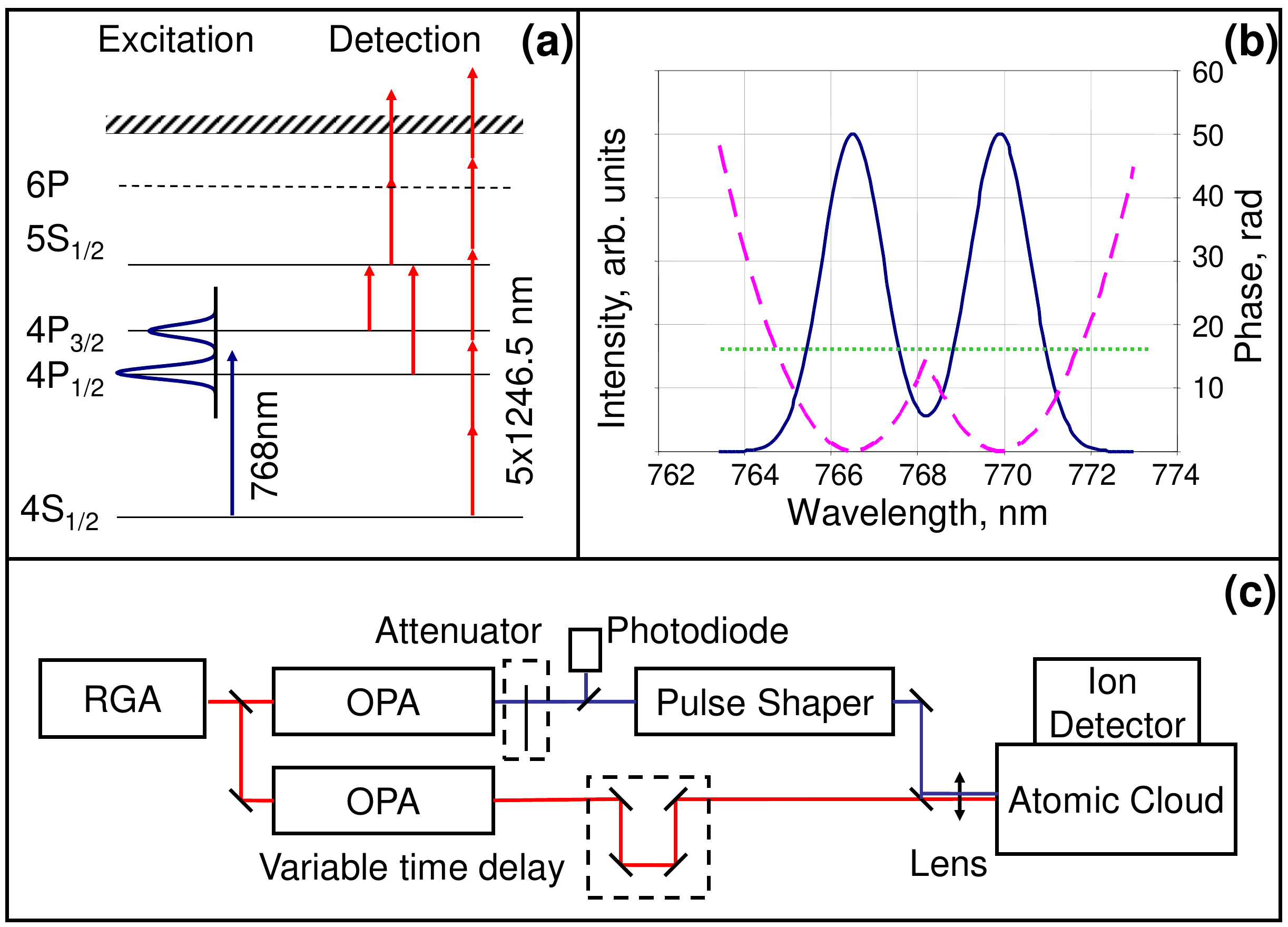}
\caption{(Color online). (\textbf{a}) Relevant quantum states of Potassium. Blue double-peak curve represents the shaped spectrum of pump pulses, whereas red arrows depict possible channels of ionizing the atoms with probe pulses.  (\textbf{b}) Spectral shaping of pump pulses with local frequency chirping around two electronic resonances at 766.5 and 766.9 nm: intensity(solid black) and phase (dashed red). Solid green line represents flat spectral phase used for a non-adiabatic excitation scheme. (\textbf{c}) Experimental setup. Two Optical Parametric Amplifiers (OPA) are pumped by a Ti:Sapphire regenerative amplifier (RGA) producing 130 fs 2 mJ pulses at 1 KHz repetition rate. One OPA is used to generate pump pulses of variable energy, controlled  by an attenuator and measured by a fast photodiode. Pump pulses are shaped with a Pulse Shaper and are weakly focused on a cloud of Potassium atoms inside a vacuum chamber. Probe pulses are delayed by a variable time delay and tightly focused on the atomic cloud. Potassium atoms, ionized by probe pulses, are accelerated towards and detected by a multichannel plate based Ion Detector.}
\label{Fig-Setup}
\end{figure}

\section{Experimental setup and detection method}
The relevant levels of Potassium atom are shown in Fig.\ref{Fig-Setup} (a). Two fine-structure levels, $4P_{1/2}$ and $4P_{3/2}$, make up the target wavepacket, which is populated from the initial ground state $4S_{1/2}$. The $D_{1}$ ($4S_{1/2}\rightarrow4P_{1/2}$) and $D_{2}$ ($4S_{1/2}\rightarrow4P_{3/2}$) transitions at 769.9 and 766.5 nm, respectively, are excited by a single broadband laser pulse of 9.5 nm full width at half maximum (\textsc{fwhm}) centered at 768.2 nm. The latter (``pump'') pulse is produced by a traveling wave optical parametric amplifier (Fig.\ref{Fig-Setup}(c)), pumped by a Ti:Sapphire femtosecond regenerative amplifier. The energy of pump pulses are attenuated and recorded with a fast photodiode prior to shaping. Hereafter, the reported pulse energies always correspond to the energy of unshaped pulses.

To apply spectral shaping, we use a home-made pulse shaper implemented in $4f$ geometry and based on a double-layer liquid-crystal spatial light modulator (SLM) \cite{Weiner2000}. The shaper controls both the phase and the amplitude of a pulse with spectral resolution of 0.14 nm. The amplitude is shaped by applying two Gaussian windows of variable relative height and width, each centered at one of the resonant frequencies (Fig.\ref{Fig-Setup}(b)). Blocking of the non-resonant spectral components is important for achieving control over the target wavepacket, as discussed below. As shown in the figure, quadratic spectral phase is added to each Gaussian window, producing variable linear frequency chirp around the corresponding resonance. Prior to applying any spectral shape described below, the phase of pump pulses is flattened using the technique of multi-photon inter-pulse interference phase scanning (MIIPS \cite{Lozovoy2006}). Flat spectral phase is also used to drive Rabi oscillations as described below. The shaped pump pulses are then focused on a cloud of Potassium atoms evaporated from a Potassium dispenser inside a vacuum chamber.

The population of the excited coherent superposition state is detected by photo-ionizing the atoms with a second (``probe'') pulse. In order to probe the region of uniform pump intensity, the probe beam is focused much tighter than the pump pulse (beam diameters of 135 and 460 $\mu$m, respectively). Our detection method is based on the weak field bichromatic coherent control scheme \cite{ShapiroBrumer2003}. The $4P_{1/2}$ and $4P_{3/2}$ states are coupled to $5S_{1/2}$ with a weak (less than 0.3 $\mu $J) femtosecond pulse (Fig.\ref{Fig-Setup} (a)). The central wavelength of probe pulses is tuned in such a way as to drive both transitions at 1243 and 1252 nm with equal efficiency. Two more photons from the same probe pulse ionized the atoms. At such low probe energy, 5-photon ionization from the ground electronic state is negligible. Extracted from the interaction regions with a series of voltage plates, the ions are identified by their time of flight and detected by a micro channel plate detector (MCP).

In order to understand the bichromatic control detection we write the wavefunction of the system as:
\begin{eqnarray}
\label{Eq-WaveFunction}
\Psi(t)&=& \sum_{i} b_{i}(t) \psi_{i} \\
&\equiv& b_{0}(t)|4S_{1/2}\rangle+b_{1}(t)|4P_{1/2}\rangle+b_{2}(t)|4P_{3/2}
\rangle. \nonumber
\end{eqnarray}
Because of the low probe energy, the ionization probability is proportional to the population of the $5S_{1/2}$ state. In the perturbative regime of interaction, this probability can be calculated as \cite{ShapiroBrumer2003}:
\begin{eqnarray}
\label{Eq-Signal}
 P(t)&\propto&\ |b_{1}(t)|^2|\epsilon(\omega_{1})|^2d_{11}
+|b_{2}(t)|^2|\epsilon(\omega_{2})|^2d_{22} + \nonumber \\
  &&2Re[b_{1}(t)b_{2}^*(t)\epsilon(\omega_{1})\epsilon^*(\omega_{2})d_{12}]
\end{eqnarray}
where $b_{1}(t)$ and $b_{2}(t)$ are the excited eigenstate amplitudes (Eq.\ref{Eq-WaveFunction}), $\epsilon(\omega_{1})$ and $\epsilon(\omega_{2})$ are \textit{probe} field amplitudes at the resonant probe transition frequencies of $4P_{1/2}\rightarrow 5S_{1/2}$ and $4P_{3/2}\rightarrow 5S_{1/2}$, respectively, and $d_{ij}=\langle\psi_{j}|\hat{d}|5S_{1/2}\rangle\langle5S_{1/2}|\hat{d}| \psi_{i}\rangle$ with $\hat{d}$ being the dipole moment operator.

The last term in Eq.(\ref{Eq-Signal}) is the interference term which depends on the difference between the relative phase of the excited wavefunctions $b_{1,2}$ and the relative phase of the two resonant probe field components $\epsilon (\omega _{1,2})$. Since the latter is constant, the ionization probability, and therefore the ion signal recorded as a function of the pump-probe time delay, oscillates as $\sin[(\omega _{1}-\omega _{2})t+\phi_{12}]$, where $\phi_{12}=\arg\{b_1^*b_2d_{12}\}$. These oscillations indicate that two $4P$ states are populated coherently, whereas their relative amplitudes and phases can be extracted from the oscillation contrast and phase, respectively \cite{Zamith2000}. In the experimental plots shown below, each data point is an average of 200 measurements.

\section{Results}
\begin{figure}
\centering
\includegraphics[width=1\columnwidth]{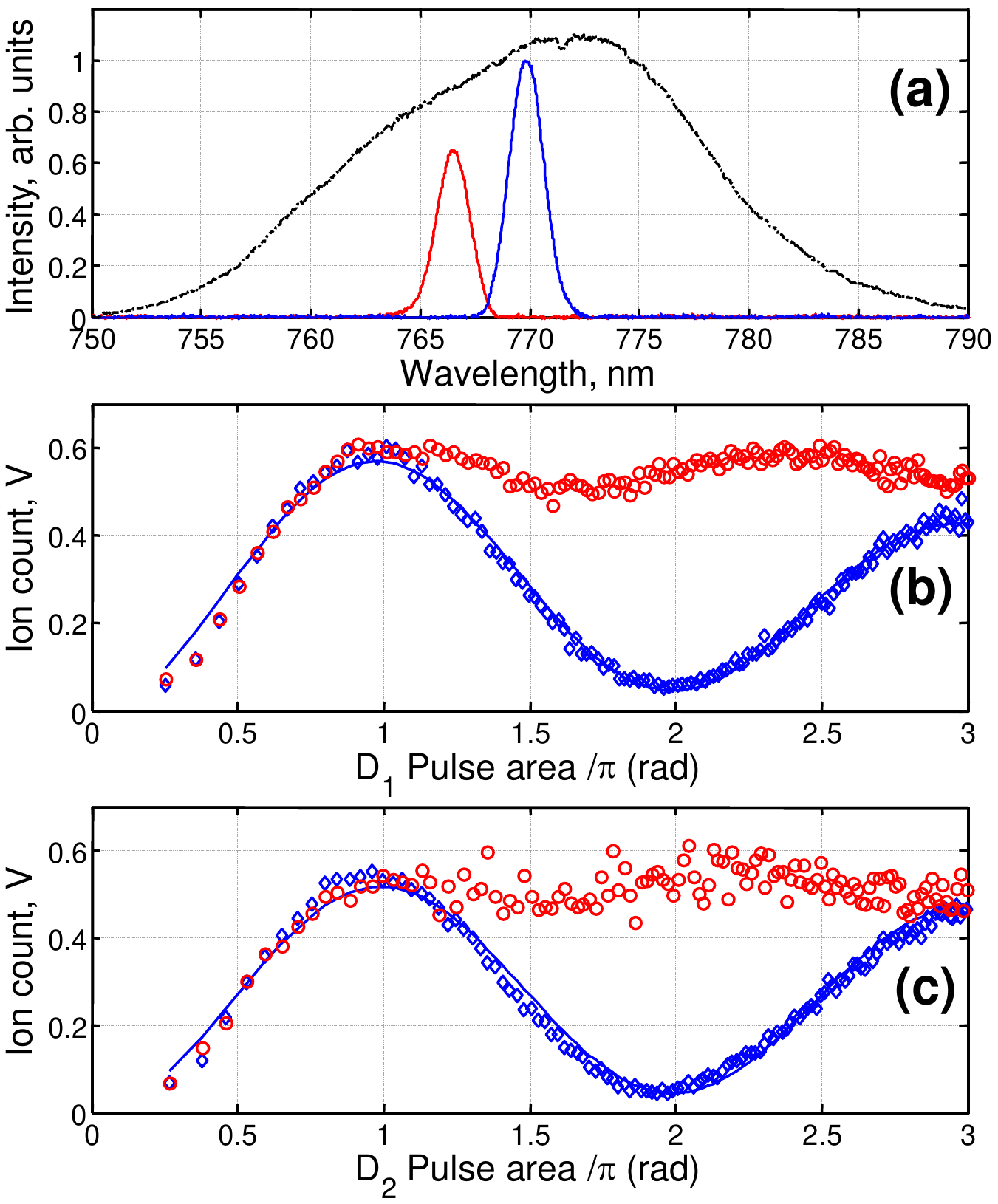}
    \caption{(Color online). Adiabatic and non-adiabatic excitation of a single ($4P_{1/2}$ or $4P_{3/2}$) excited state. \textbf{(a)} Pump pulse spectra used to excite either $D_{1}$ or $D_{2}$ transitions separately (blue and red solid lines, respectively). Spectral profile of the initial unshaped femtosecond pulse is shown as dashed black curve. \textbf{(b, c)} Non-adiabatic (blue diamonds) and adiabatic (red circles) population transfer into $4P_{1/2}$ (b) and $4P_{3/2}$ (c) states. Ion signal, proportional to the target state population, is plotted as a function of the corresponding pulse area, $A_{1,2}$. Solid lines show the anticipated $\sin^{2}(A_{1,2})$ dependence fitted to the experimental data.}
\label{Fig-Results1D}
\end{figure}

We first execute population transfer into each of the $4P$ states separately. Non-adiabatic interaction results in familiar Rabi oscillations which serve as a convenient tool for calibrating both the excitation pulse area and excited state population. The original spectrum and the two Gaussian windows of 1.8 nm FWHM used for amplitude shaping are shown in Fig.\ref{Fig-Results1D}(a). To drive Rabi oscillations on a selected transition, a single Gaussian filter was applied around the resonant frequency of $D_{1}$ or $D_{2}$ transition. Hereafter, we refer to the corresponding narrow-band pulses as ``$D_{1}$ pulse'' and ``$D_{2}$ pulse'', respectively. In both cases, the spectral phase was kept flat across the whole spectrum of a pulse. Rabi oscillations are shown in Fig.\ref{Fig-Results1D} for both $D_{1}$ (b) and $D_{2}$ (c) lines. Given the available laser power, we were able to reach pulse areas of up to 3$\pi$ on each transition. By fitting the data with a $\sin^{2}A_{1,2}$ function, we calibrate the areas of $D_{1}$ and $D_{2}$ pulses ($A_{1}$ and $A_{2}$, respectively) versus their energy. Simultaneously, we calibrate the magnitude of the ion signal corresponding to the pulse area of $\pi$. These calibrations are later used for assessing the completeness of the population transfer.

In order to execute an adiabatic passage into a single excited state, a frequency chirp is introduced by applying quadratic spectral phase shaping around one resonant frequency $\omega _{1,2}$, i.e. $\varphi(\omega )=\frac{\alpha}{2}(\omega -\omega _{1,2})^2$ (Fig.\ref{Fig-Setup}(b)), while blocking the Gaussian window around the other resonance.  The frequency chirp is gradually increased by the shaper until the population transfer shows AP-like saturation with intensity, such as that shown in Figs.\ref{Fig-Results1D}(b,c) by open circles for both $D_{1}$ and $D_{2}$ transitions. Independent measurement of the applied chirp yields $\alpha=270\times 10^3$ fs$^{2}$. For both transitions, the ion signal reaches its maximum at a pulse area of $\approx \pi $ and stays relatively flat with increasing pulse energy. The maximum in the AP efficiency is seen to coincides with a maximal number of Rabi oscillations, attesting to nearly complete population transfer. The residual oscillations, reproduced in our numerical simulations (not shown), are due to the pixelization of the spatial light modulator.

Once the pulse area and chirp required to satisfy adiabaticity conditions are determined, we can execute complete population transfer into a superposition of $4P_{1/2}$ and $4P_{3/2}$ states. If both states are populated coherently, quantum beats should be observed in the ionization signal as a function of the probe pulse delay  according to the interference term in Eq.(\ref{Eq-Signal}). In both the non-adiabatic and adiabatic approaches, the amplitude of the spectrum of the pump pulses is shaped by opening both the $D_{1}$ and $D_{2}$ Gaussian windows \textit{simultaneously}. Interference of the $D_{1}$ and $D_{2}$ pulses in the time domain results in a train of pulses separated by a period of the quantum evolution of the wavepacket. The latter is inversely proportional to the fine-structure splitting of $4P$ state (1.73 THz) and equals 578 fs. A numerical example of such pulse train is shown in Fig.\ref{Fig-APSimulations}(b).
\begin{figure}
\centering
\includegraphics[width=1\columnwidth]{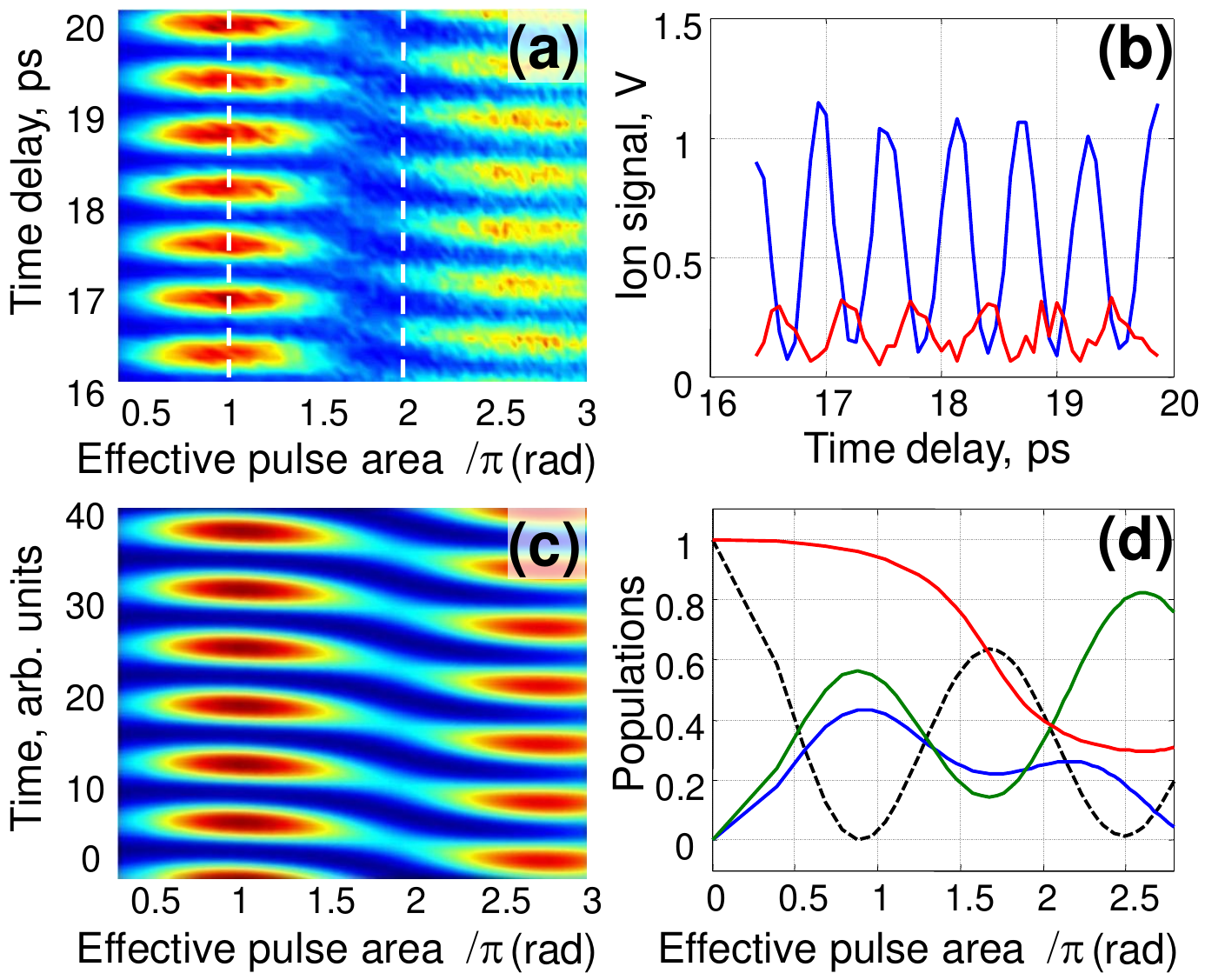}
\caption{(Color online). Non-adiabatic population transfer into a superposition of $4P_{1/2}$ and $4P_{3/2}$ states. The two-dimensional plots (\textbf{a}: experiment, \textbf{c}: calculation) show the ion signal (color coded) as a function of the effective pulse area and pump-probe time delay. Oscillations along the vertical (time) axis reflect quantum beating due to the time evolution of the wavepacket. The quantum beats are shown in (\textbf{b}) for pulse areas of $\pi$ (blue) and $2\pi$ (red), corresponding to the dashed vertical white lines in (\textbf{a}). In plot (\textbf{d}), we present the calculated populations of the two excited states, $4P_{1/2}$ (blue) and $4P_{3/2}$ (green), and the relative phase (red) between the corresponding wavefunctions, as a function of the effective pulse area of the excitation field. The ground state population is shown as a dashed black line.}
\label{Fig-Results2DRabi}
\end{figure}

First, the spectral phase of pump pulses is kept flat across the whole double-peak spectrum. In the time domain, flat spectral phase translates into a train of pulses with constant carrier oscillation frequency and no extra phase shift between consecutive pulses \cite{Shapiro2009}. The resulting Rabi oscillations between the ground state and the excited wavepacket are shown in Fig.\ref{Fig-Results2DRabi}(a). The two-dimensional plot shows the ion signal as a function of the time delay (vertical axis) and effective pulse area (horizontal axis), which in the case of multiple excitation channels can be used as a convenient scale of the interaction strength. It is defined as:
\begin{equation}
A_{\textmd{eff}}=\sqrt{A_{1}^2+A_{2}^2}
\label{Eq-EffArea}
\end{equation}
where $A_{i} \equiv \int_{-\infty}^{\infty} \Omega _{i}(t) \text{d}t $, with $\Omega_{i}$ being the time dependent Rabi frequencies for the $i$-th transition.

Oscillations along the vertical axis indicate quantum beating between $4P_{1/2}$ and $4P_{3/2}$ states. Both the experimental and numerical plots (Fig.\ref{Fig-Results2DRabi}(a,c)) demonstrate strong dependence of the result of the excitation on the effective pulse area. Rabi oscillations between the ground state and the excited wavepacket are manifested by the periodic re-appearance of the beat signal. Thus, high beating contrast at $A_{\textmd{eff}}=\pi$ disappears almost completely at $A_{\textmd{eff}} \approx 2\pi $, as shown by the vertical cross-sections of the two-dimensional data (white dashed lines) in plot (b).

We note that both the contrast of the beat signal and its phase are affected by an increase in the pulse energy. We attribute this phase change to the energy-dependent AC Stark shifts of the levels, caused by the strong non-resonant components of the polychromatic excitation field (see schematic illustration in Fig.\ref{Fig-LevelDiagram}). In the case considered here, the $4P_{1/2}$ level is shifted due to the presence of an off-resonant $D_{2}$ pulse, whereas the $4P_{3/2}$ level is shifted by an off-resonant $D_{1}$ field. Since the shifts are in general unequal, the accumulated quantum phase of the two wavefunctions, $|4S_{1/2}\rangle$ and $|4S_{3/2}\rangle$, depends on the energy of both pulses. The effect is reproduced by the numerical calculations shown in Fig.\ref{Fig-Results2DRabi}(c).

For pulse areas of order $\pi $ and higher, typically needed for significant population transfers, such dynamic cross-talks between simultaneously driven transitions become substantial when the pulse bandwidth approaches the energy separation between the levels. In this case, dynamic Stark shifts cannot be neglected and multiple interaction channels cannot be treated independently. This significantly complicates the non-adiabatic dynamics, making controlled and complete population transfer hard to achieve. Fig.\ref{Fig-Results2DRabi}(d) demonstrates the degree to which the dependence of the $|b_{1}(t)|^2$ and $|b_{2}(t)|^2$ populations on pulse energy deviates from the periodic Rabi oscillations behavior, once the effective pulse area exceeds $\pi $.

\begin{figure}
\centering
\includegraphics[width=1\columnwidth]{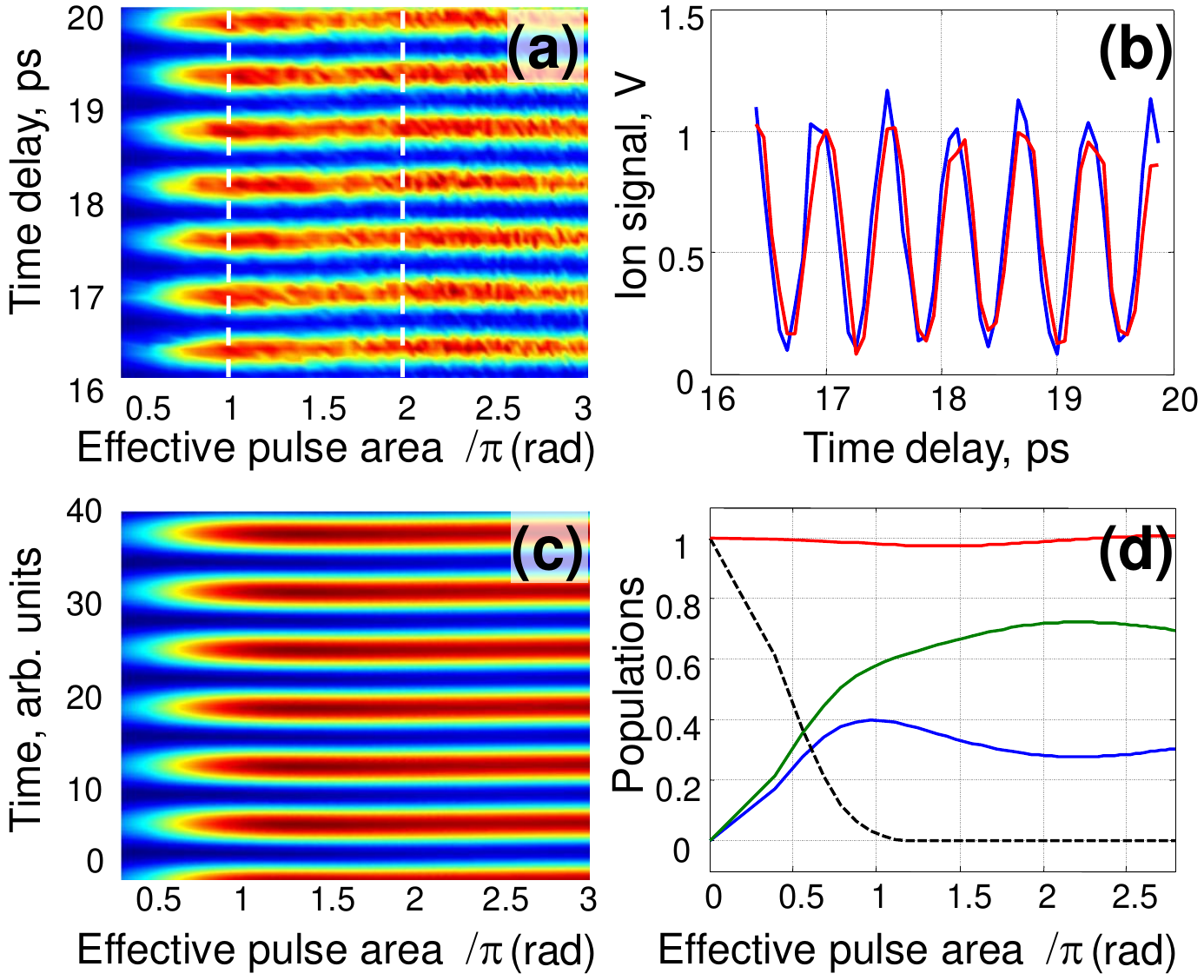}
\caption{(Color online). Adiabatic population transfer into a superposition of $4P_{1/2}$ and $4P_{3/2}$ states. The two-dimensional plots (\textbf{a}: experiment, \textbf{c}: calculation) show the ion signal (color coded) as a function of the effective pulse area and pump-probe time delay. The oscillations along the vertical (time) axis reflect quantum beating due to the time evolution of the wavepacket. The quantum beats are shown in (\textbf{b}) for pulse areas of $\pi$ (blue) and $2 \pi$ (red), corresponding to the dashed vertical white lines in (\textbf{a}). In plot (\textbf{d}), we present the calculated populations of the two excited states, $4P_{1/2}$ (blue) and $4P_{3/2}$ (green), and the relative phase (red) between the corresponding wavefunctions, as a function of the effective pulse area of the excitation field. The ground state population is shown as a dashed black line.}
\label{Fig-Results2DAP}
\end{figure}
In striking contrast to Rabi flopping, when local chirp is added to both the $D_{1}$ and $D_{2}$ spectral windows (Fig. \ref{Fig-Setup}(b)), the observed quantum beats become insensitive to the pulse area (Fig.\ref{Fig-Results2DAP}(a,c)). Using our experimental observation of separate adiabatic passages into each level (Fig.\ref{Fig-Results1D}), we set the frequency chirp to $270\times 10^3$ fs$^2$, for both the $D_{1}$ and $D_{2}$ pulses. As seen in Fig.\ref{Fig-Results2DAP}(b), the contrast of the experimentally observed quantum beating changes only little throughout the wide range of pulse areas, $A_{\textmd{eff}} \approx \pi $ to $A_{\textmd{eff}} \approx 3\pi $. This demonstrates the stability of the population transfer against pulse energy in agreement with the adiabatic passage scenario.

Though clearly more stable than in the non-adiabatic regime, the beat signal gives only indirect evidence of the measure of the robustness of the population transfer. Unfortunately, extracting the absolute values of the population transferred to $4P_{1/2}$ and $4P_{3/2}$ from the measured beat signal proved inaccurate. We therefore calculate these populations numerically. The results, presented in Fig.\ref{Fig-Results2DAP}(d), show that the population ratio varies with the excitation pulse energy. A number of simple approaches to controlling this ratio are discussed later in the text. For the time being we note that an increase in the pulse area leaves the phase of the excited wavepacket intact. This statement is confirmed experimentally and theoretically, as shown in panels (b) and (d), respectively.

\begin{figure}
\centering
\includegraphics[width=1\columnwidth]{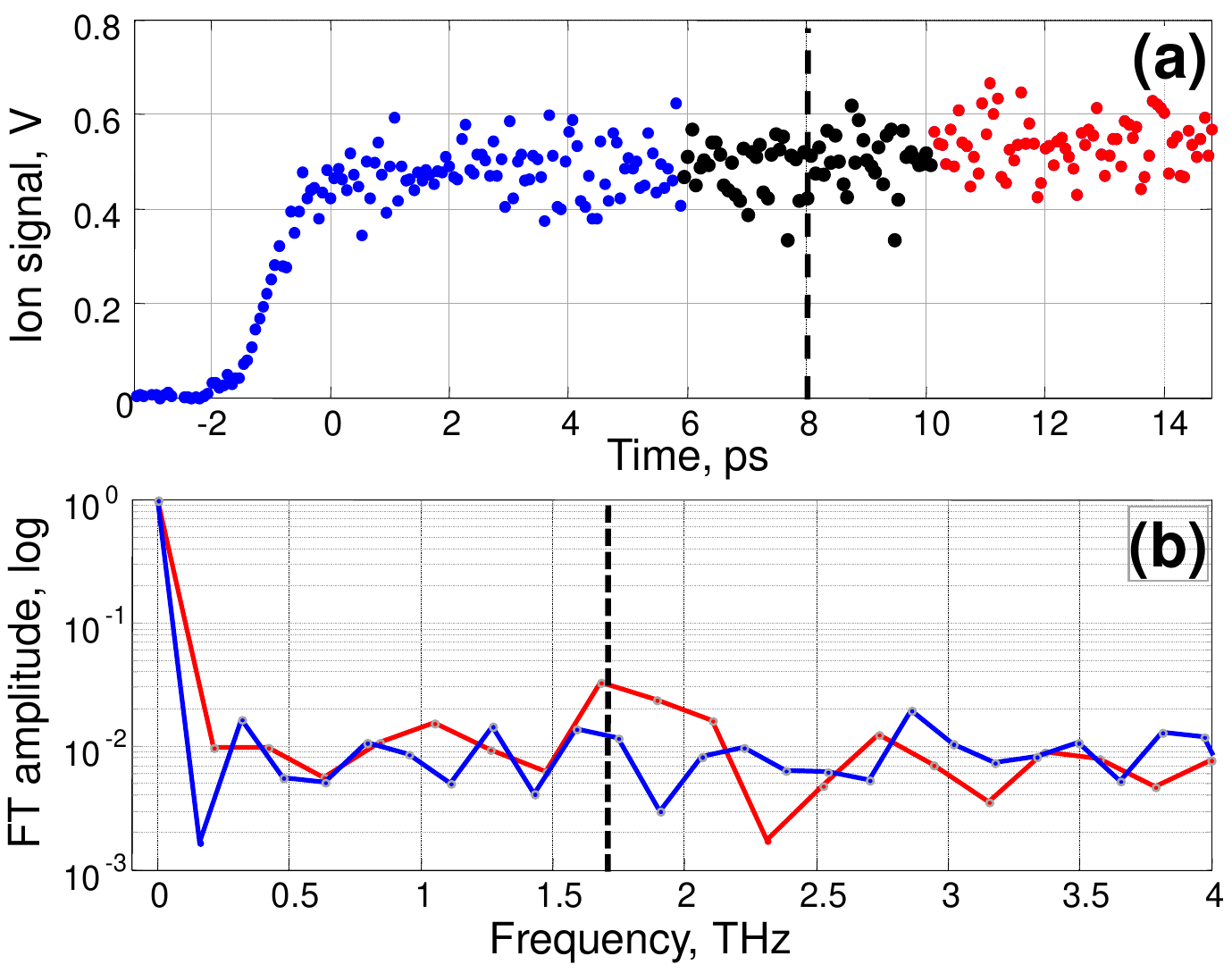}
\caption{(Color online). Demonstration of the population transfer completeness. (\textbf{a}) Ion signal as a function of the pump-probe time delay for a sequence of frequency chirped $D_{1}$ and $D_{2}$ pulses separated by 8 ps (marked by dashed vertical line). Blue (red) dots correspond to the signal before (after) the arrival of the second pulse. (\textbf{b}) Fourier transform of the ion signal before (blue) and after (red) the arrival of the second pulse (log scale). Vertical dashed line marks the frequency of the quantum beating (1.73 THz) where a strong peak is expected in the case when both states are populated.}
\label{Fig-CompletnessResult}
\end{figure}
In contrast to the non-adiabatic dynamics, AP exhibits a threshold at $A_{\textmd{eff}} \approx \pi $, beyond which the ground level remains empty, even though the population ratio between the excited levels may vary. This stability of the completeness of population transfer against variations in pulse energy is best demonstrated by comparing the numerical results of Figs.\ref{Fig-Results2DRabi}(d) and \ref{Fig-Results2DAP}(d). It also forms the basis for the proposed schemes of controlling the population distribution among excited states. We therefore seek an independent experimental confirmation of the completeness of population transfer.

In the experiments described above, we have used Rabi oscillations between the ground state and one of $4P$ excited states for calibrating the population transfer efficiency, assuming that the first maximum of these oscillations corresponds to a complete transfer. However, even though a $\pi$-pulse is expected to drive the whole population to the excited state, it is conceivable that completeness is not fulfilled due to the sensitivity of the latter to frequency detuning. In order to perform an independent check of the completeness of the population transfer, we have carried out the following experiment: We delay the $D_{1}$ pulse by approximately 8 ps with respect to the $D_{2}$ pulse by inserting a thin piece of glass in front of the corresponding spectral window in the Fourier plane of the pulse shaper. If the $D_{2}$ pulse, arriving first, drives all the atoms to $P_{3/2}$ state and depletes the ground state, the $D_{1}$ pulse, arriving second, will leave the system unchanged. As all atoms are residing in a single $P_{3/2}$ state, no quantum beats would be observed. Alternatively, if the excitation driven by the first pulse is not complete, the second pump pulse would move the remaining ground state population to $P_{1/2}$ state, resulting in the appearance of quantum beating.

To verify this, we set the areas of both $D_{1}$ and $D_{2}$ pulses (now separated in time) to $\pi $, and their frequency chirp to $270 \times 10^3$ fs$^{2}$. As can be seen in Figs.\ref{Fig-Results1D}(b,c), these parameters are sufficient for reaching maximum population transfer to either $P_{1/2}$ or $P_{3/2}$ when the corresponding pulse is acting alone. The results of the interaction with a sequence of two pulses are presented in Fig.\ref{Fig-CompletnessResult}. The first pulse arrives at approximately 0 ps, while the time of arrival of the second pulse is about 8 ps. Fourier analysis of the ion signal after the arrival of the second pulse (Fig.\ref{Fig-CompletnessResult}(c)) shows essentially no quantum beating, ensuring that the population of $4P_{1/2}$ state is below 3\%.

\begin{figure}[b]
\centering
\includegraphics[width=1\columnwidth]{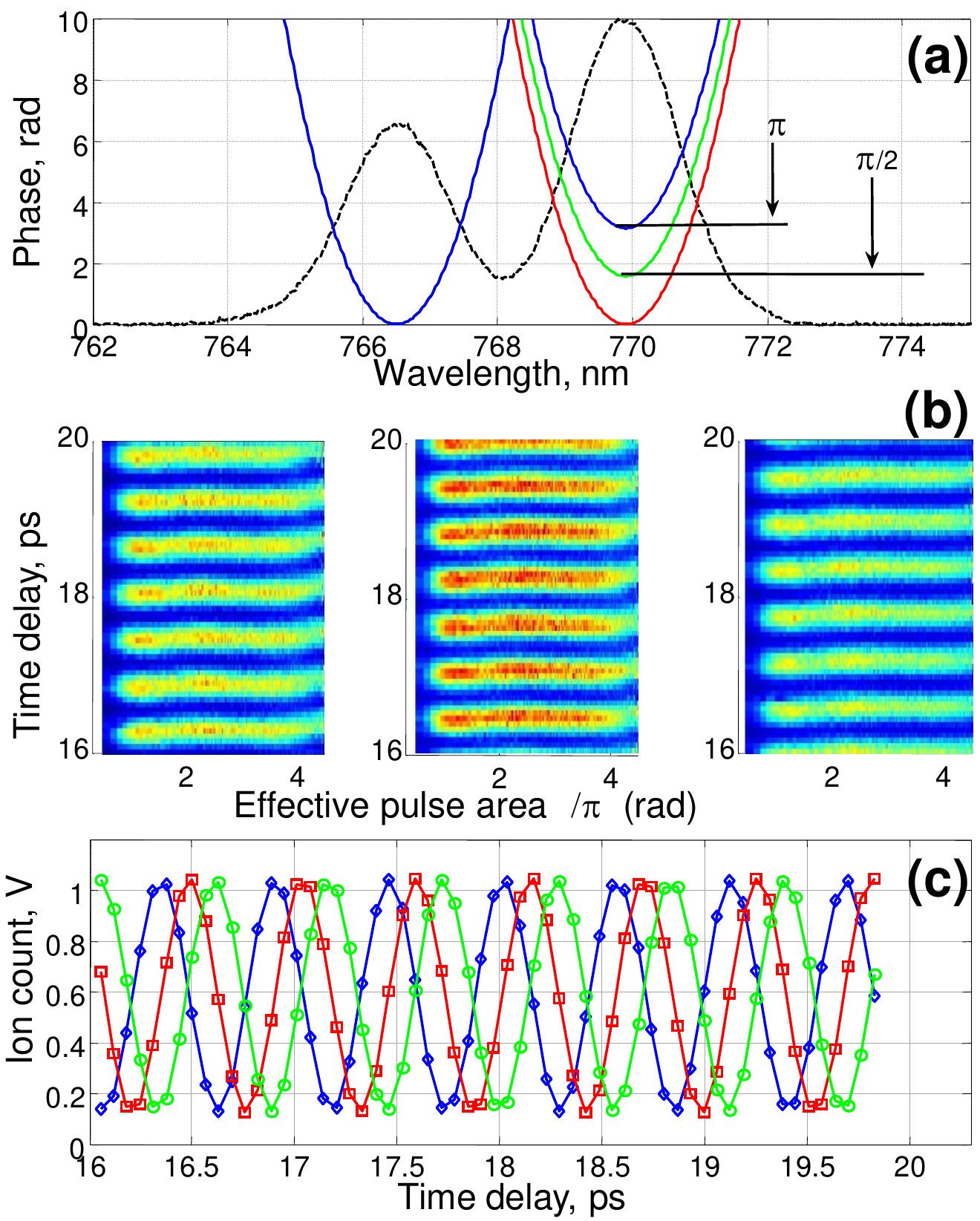}
\caption{(Color online). Experimental control over the quantum phase of an excited wavepacket. \textbf{(a)} To control the relative phase of the eigenstates in the target superposition state, an extra constant phase shift of 0 (red), $\pi/2$ (green) and $\pi$ radian (blue) is added to the spectral phase of $D_{1}$ pulses. Black dashed line shows the double-peaked spectral amplitude of the pulse. The introduced phase shift results in a corresponding vertical shift of the detected quantum beats shown in \textbf{(b)}. \textbf{(c)} Vertical cross-sections of the two-dimensional scans in (\textbf{b}) at pulse area of $\pi$.}
\label{Fig-PhaseControl}
\end{figure}

As pointed out by us in the past \cite{Shapiro2009}, piecewise adiabatic passage potentially combines the efficiency and robustness of AP with the ability to excite complex superposition states (e.g. wavepackets) and control the makeup of the excited state wavefunction \cite{Shapiro2009}. Control over the phase of a spin-orbit wavepacket in Potassium has been previously demonstrated experimentally in the weak field regime \cite{Chatel2008}. Here we show that the relative phase between the two eigenstates of the target superposition can be controlled even when the field is strong enough to ensure the adiabaticity of process. Control over the phase is implemented by adding an extra constant phase to one of the pulses, $D_{1}$ or $D_{2}$, on top of the local frequency chirp in the respective spectral window. The resulting phase profile, attained using the pulse shaper, is shown in Fig.\ref{Fig-PhaseControl}(a) for a relative phase of 0, $\pi/2$ and $\pi$ radian. Two-dimensional energy-time scans for these three phase shifts are shown in panel (b). The vertical cross-sections of each two-dimensional plot, displaying quantum beat patterns corresponding to $\approx\pi$ pulses, are plotted in panel (c). The  phase of the oscillations, which directly reflects the relative phase of $P_{1/2}$ and $P_{3/2}$ eigenstates, closely follows the extra phase shift introduced via the pulse shaper.

\begin{figure}[t]
\centering
\includegraphics[width=1\columnwidth]{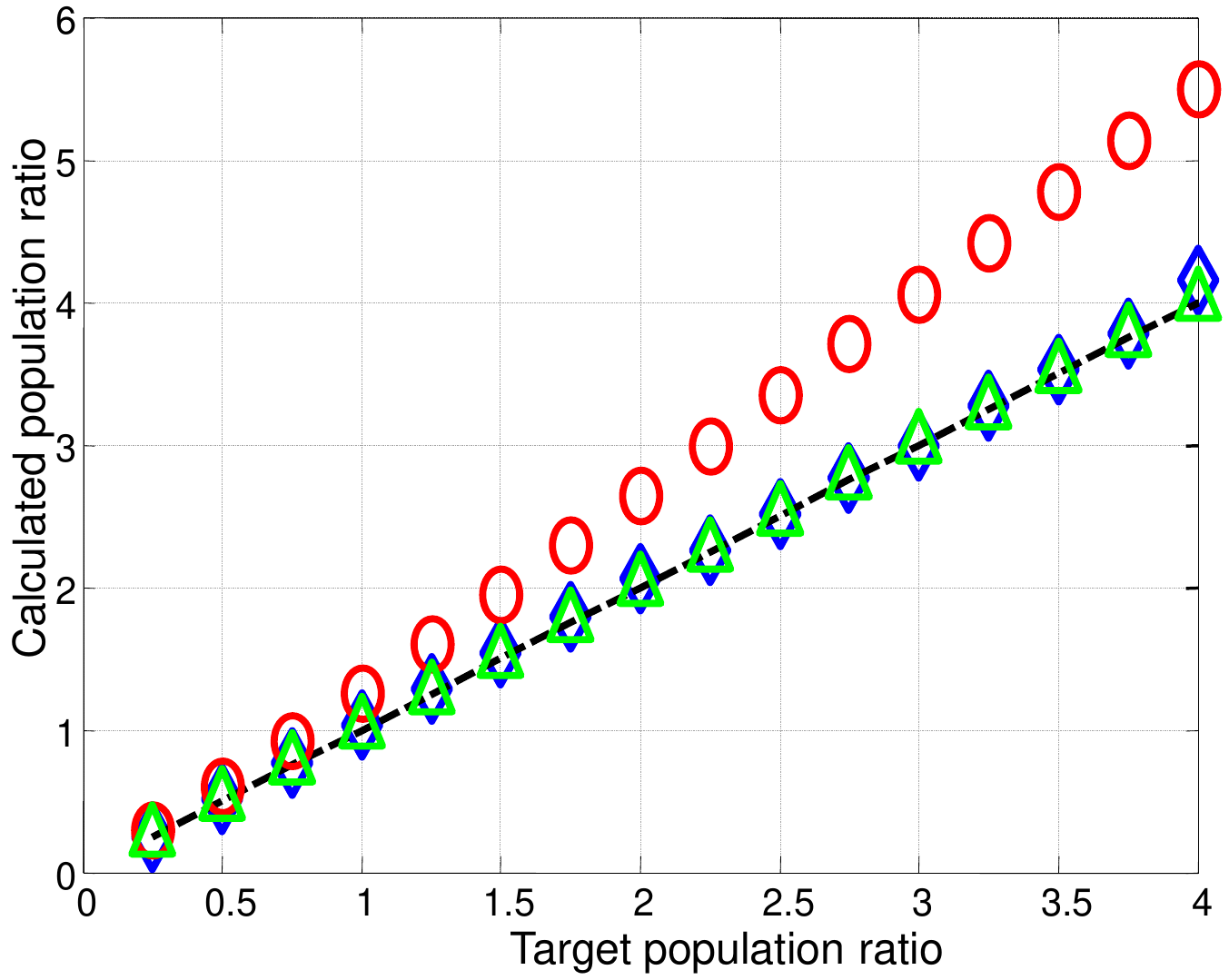}
\caption{(Color online). Quality of population control with piecewise adiabatic passage. Numerical simulations demonstrate possible discrepancy between the achieved population ratio $\beta $ of two $4P$ states (vertical axis)  and its target value (horizontal axis), for different control methods (see text for symbol description).}
\label{Fig-AmplitudeControl}
\end{figure}
One may expect that the amplitudes of the eigenstates in the target superposition can be controlled individually, similarly to the phase control described above. As was shown in \cite{Shapiro2009} for the case of \textit{negligible Stark shifts}, the state amplitudes at the end of the interaction are simply proportional to Rabi frequencies of the corresponding transitions. Thus, the distribution of populations among excited states (here, $4P_{1/2}$ and $4P_{3/2}$) can be controlled by changing the relative strength of the corresponding spectral components of the excitation field (here, the energies of $D_{1}$ and $D_{2}$ pulses). This simple strategy fails, however, once the dynamic Stark shifts become comparable to the energy bandwidth of the laser pulses. This situation is illustrated in Fig.\ref{Fig-LevelDiagram}(b), where one can see that unequal Stark shifts effectively change the time at which each adiabatic passage is executed. The farther these APs from being simultaneous, the bigger the deviation of the population distribution from the expected one.

Poor accuracy in retrieving state populations from the observed beat signal does not allow us to investigate the possibilities of amplitude control experimentally. Here, we present numerical analysis of various ways of achieving reasonably high degree of control over the excited state populations by either avoiding, or compensating for the detrimental effects of the Stark shifts. The results are summarized in Fig.\ref{Fig-AmplitudeControl}, in which the calculated ratio $\beta \equiv \left|b_1\right|^2 / \left|b_2\right|^2$ is plotted versus its target value for different control methods. In all calculations, the effective pulse area was set to $\pi$, and the local spectral chirps near $D_1$ and $D_2$ lines was equal to $270 \times 10^3$ fs$^2$.

We first note that for the parameters used in our experiments, calculated values of $\beta $ (red circles) lie within 37.5\% of those expected from a simple model which does not take Stark shifts into account (diagonal dashed line). Assuming the correct phases of the state amplitudes, this corresponds to the transfer fidelity (projection on the target wavepacket) varying between 1 and 0.997. The discrepancy can be further reduced by narrowing the spectral bandwidth of the excitation pulses. Decreasing it from 1.8 nm (as used here) to 0.18 nm, dramatically improves the quality of amplitude control, marked as blue diamonds. It is important to note that such improvement comes at the expense of much longer (by a factor of 10) interaction time, and higher spectral resolution of a pulse shaper, which will be necessary for the frequency chirping of a spectrally narrower pulse.

Unlike the case of independent adiabatic passages, i.e. when Stark shifts can be ignored, the Stark-induced dynamical cross-talk between different AP channels makes $\beta $ sensitive not only to the ratio of $D_{1}$ and $D_{2}$ pulse energies, but also to their sum. This certainly reduces the robustness of the proposed method of amplitude control,though not the robustness of PAP itself. In other words, in PAP, the population transfer remains complete, even though the shape of the excited wavepacket may change in response to changes in the overall pulse energy. Utilizing this important property of piecewise adiabatic passage, we suggest a simple adaptive strategy of controlling the eigenstate amplitudes in the excited superposition state. By accurately measuring populations of the excited states (not available in our current experimental setup), a correction can be introduced into the relative energy of each spectral component of the excitation field, proportional to the deviation of the observed population distribution from its target shape. The numerical results of such an iterative procedure appear as green triangles in Fig.\ref{Fig-AmplitudeControl}. After only two iterations, the calculated ratio between $\left|b_1\right|^2$ and $\left|b_2\right|^2$ is brought within 0.2\% of its target value! Unlike the more general schemes of adaptive control \cite{Judson92}, the present technique involves only as many control variables as there are excited states (in our case, two), and converges very quickly due to the inherent robustness of AP.

\section{Conclusion}
We have presented experimental and numerical studies of the population transfer using piecewise adiabatic passage from a single ground state to a superposition of excited states. As in conventional adiabatic population transfer into a single state, our method of piecewise adiabatic passage into a wavepacket is insensitive to the driving field amplitude. The method allows for complete population transfer and offers control of both the phase and amplitudes of the eigenstates composing the target superposition state. The latter are retrieved by applying bichromatic control to the process of photoionization.

\begin{acknowledgements}
This work has been supported by the CFI, BCKDF, NSERC and DTRA.
\end{acknowledgements}


\end{document}